\documentstyle[11pt,epsfig,psfig]{article}
\def\date
   {\noindent Date: \today \par
    \medskip}

\renewcommand{\omega}{w}

\begin{document}
%----------------------------------------------------------------------
% Title page:
%----------------------------------------------------------------------
\begin{center}

 {\Large \bf Coarsening of a Class of Driven Striped Structures}

\end{center}
\vspace*{2ex}
\begin{center}
\normalsize  M. R. Evans$^1$, Y. Kafri$^2$, E. Levine$^2$, and D. Mukamel$^2$
\\[2ex] $^1${ \it Department of Physics and Astronomy,
University of Edinburgh, Mayfield Road, Edinburgh EH9 3JZ, United Kingdom.
\\[1ex]
 $^2${ \it Department of Physics of Complex
Systems, The Weizmann Institute of Science, Rehovot 76100, Israel. }
\\[4em] }
\end{center}

\noindent {\bf Abstract:} The coarsening process in a class of
driven systems exhibiting striped structures is studied. The
dynamics is governed by the motion of the driven interfaces
between the stripes. When two interfaces meet they coalesce thus
giving rise to a coarsening process in which $\ell(t)$, the
average width of a stripe, grows with time.  This is a
generalization of the reaction-diffusion process $A+A{\to }A$ to
the case of extended coalescing objects, namely, the interfaces.
Scaling arguments which relate the coarsening process to the
evolution of a single driven interface are given, yielding growth
laws for $\ell(t)$, for both short and long time. We introduce a
simple microscopic model for this process. Numerical simulations
of the model confirm the scaling picture and growth laws. The
results are compared to the case where the stripes are not driven
and different growth laws arise.\\[1em]
%

%\rule{7cm}{0.2mm}
\date
\begin{flushleft}
\parbox[t]{12.5cm}{ }
\\[2mm]
\parbox[t]{3.5cm}{\bf PACS numbers:} 05.40.-a, 05.70.Ln, 05.70.Np, 64.60.Cn
\\[2mm]
\end{flushleft}
\normalsize \thispagestyle{empty} \mbox{} \pagestyle{plain}
\newpage
\setcounter{page}{1} \pagestyle{plain} \setcounter{equation}{0}
%%%%%%%%%%%%%%%%%%%%%%%%%%%%%%%%%%%%%%%%%%%%%%%%%%%%%%%%%%%%%%%%%%%%%%%%%%%%%%%

\section{Introduction}
Coarsening processes have been extensively studied both
experimentally and theoretically over the past decades
\cite{GSS,BRAY1}. Most of these studies deal with the way a system
approaches its {\it thermal equilibrium} state. For example, when
a liquid is quenched to  temperatures below the liquid-gas
transition point a coarsening process takes place as the system
evolves towards the equilibrium phase separated state. In these
systems the order parameter (the density) is conserved and the
evolution proceeds by either nucleation growth or by spinodal
decomposition \cite{CH}. Similar processes take place in magnetic
systems where a system
evolves towards its magnetically ordered state
when quenched below its critical point. In that case, however, the
order parameter (the magnetization) need not be conserved by the
dynamics and the details of the coarsening mechanism may differ
from that of conserving dynamics.

Typically coarsening processes are characterized by a single
growing length scale $\ell (t)$, for example the average domain size in
the system. In many cases, at late times the system reaches a
scaling regime where $\ell(t) \sim t^n$. The value of the exponent
$n$ usually depends on the symmetry of the system and its
conservation laws. It has been found that for a scalar order
parameter that is not conserved under the dynamics (as is the case
in many magnetic systems) the growth exponent is $n=1/2$ \cite{AC}. This has
been demonstrated theoretically by numerous studies of Ising-type
models with Glauber dynamics \cite{Glauber}. On the other hand,
when the order parameter is conserved (such as in liquid gas
transitions and in phase separation in binary mixtures) the
coarsening process is slower and the growth exponent was found to
be $n=1/3$. This was first demonstrated by
Lifshitz, Slyozov and Wagner \cite{LS,Wagner} and has been confirmed
by many studies of Ising models with Kawasaki dynamics \cite{HUSE,BRAY1}.

More recently attention has focused on systems {\it far from
thermal equilibrium}, driven by an external field \cite{SZ}. In many cases
such systems reach a steady state in which, unlike the equilibrium
case, detailed balance is not obeyed. The lack of detailed balance
allows for many phenomena which do not occur in thermal
equilibrium, such as phase separation and symmetry breaking in one
dimensional systems \cite{David,Brazil}. Most of the recent studies of
these systems have been focused on the properties of the steady
state itself rather than on the evolution towards it. There are
however indications that coarsening processes in these systems may
be rather different from those of systems evolving toward their
equilibrium state. For example a study of the evolution of a
driven Ising model with conserving dynamics has shown that in one
dimension the average domain size grows as $t^{1/2}$
\cite{CBD,SKR} rather than the usual $t^{1/3}$ expected for
non-driven systems.

The presence of the drive introduces a preferred direction in space
making the systems inherently anisotropic. In many of these systems
this results in striped structures. Typical examples are the stepped
structures which occur in surface growth \cite{JW} and wind ripples
formed in sand \cite{BAG,CMRV}. Many models of driven systems have
been introduced and studied in recent years. The
striped structures that naturally emerge  may be
oriented either parallel or perpendicular to the driving field,
depending on the details of the dynamics \cite{SZ}. For example in a
driven Ising model introduced by Katz {\it et al} \cite{KL} stripes
parallel to the driving field of alternating up and down spins are
found at low temperatures. When the system is quenched to the low
temperature phase at high values of the driving field narrow stripes
are formed on a short time scale. On a longer time scale a coarsening
process takes place in which the typical width of these stripes grows
in time as the system evolves towards a fully phase separated state.
This coarsening process is rather different from the one taking place
in a system evolving towards its equilibrium state and is not well
understood \cite{YRHJ,ZSS}.  On the other hand, stripes perpendicular
to the drive direction are observed in related spin-1 type models
where two types of oppositely charged particles move in the presence
of neutral vacancies \cite{SHZ,SZ2,YKMsC}.

Recently, a different class of driven models, where phase
separation takes place even in one dimension, has been introduced
\cite{EKKM1,EKKM2,LR,LR2,AR1,AR2}. In
this class of models three or more types of particle are driven in
a preferred direction under local dynamics.  The important feature
in these models is that the {\it local} dynamics results in a
phase separated state even in one dimension ($D=1$).  It has been
demonstrated that in $D=1$ the coarsening process which
accompanies the phase separation is slow, by which it is meant
that the average domain size grows only logarithmically with time.
An extension of one of the models belonging to this class to two
and higher dimensions \cite{YBEM} showed that in dimensions
greater than one stripes of alternating types of particles are
formed perpendicular to the driving field. The width of the
stripes along the direction of the driving field is found to grow
as $\log (t)$, which is the same growth law as the one dimensional
case. This is related to the fact that the interfaces which
separate adjacent stripes are macroscopically smooth.

Given the common occurrence of striped structures in different classes
of driven models, it is of interest to explore the possible coarsening
phenomena within a wider range of models exhibiting these structures.
For example, in the preceding paragraph we described a driven system
with conserving dynamics exhibiting slow coarsening. This behavior
can be attributed to the smoothness of the interfaces separating
neighboring stripes.  A natural question is as to how the coarsening
of stripes is altered when one considers more general scenarios such
as, say, non-conserving dynamics or dynamics which leads to rough
interfaces separating the stripes.

\begin{figure}
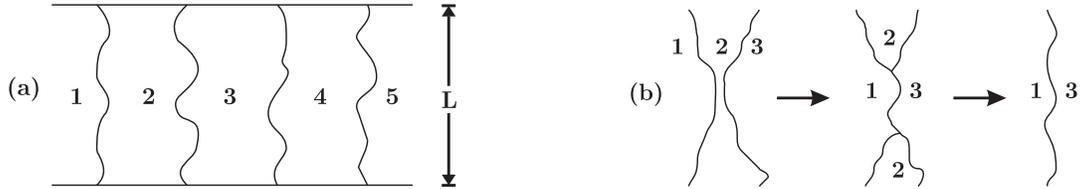

\begin{center}
\epsfxsize 6 cm \epsfbox{stripes.eps} \hspace{2.0cm} \epsfxsize 6
cm \epsfbox{merge.eps} \caption{Schematic representation of the
model. (a) Domains are aligned parallel to each other, and are
assigned one of $q \rightarrow \infty$ spin states in an ordered
manner. (b) When two interfaces meet they coalesce into a single
interface.  } \label{fig:pict}
\end{center}
\end{figure}

To investigate this issue we study in this paper a simple model
for the evolution of driven striped structures. In the model
stripes emerge oriented perpendicular to the driving field. The
microscopic dynamics is non-conserving and rough interfaces
between the stripes arise. We consider a system of infinite extent
in the direction parallel to the drive and of finite size $L$ in
the other directions (see Fig. \ref{fig:pict}(a)). A detailed
definition of the microscopic model and its dynamics is given in
Section 2. Here we just describe the salient features:
\begin{itemize}
  \item A microscopic state of the system is given by the
configurations of the interfaces separating adjacent stripes (Fig.
\ref{fig:pict}(a)). These interfaces are assumed to be single
valued, that is no bubbles or overhangs are present.
  \item Each interface evolves under local driven dynamics. For example
one can consider growth dynamics belonging to the KPZ universality class.
All interfaces evolve under the same dynamical rules, and thus, in
particular, they all move in the same direction and with the same
average velocity.
  \item  When two interfaces meet at a point, they locally merge to form
a single interface. The evolution is such that after some time the
entire two interfaces coalesce (Fig. \ref{fig:pict}(b)).
\end{itemize}

As a result of this dynamical process the number of interfaces
keeps decreasing and the average width of a stripe, $\ell(t)$,
increases with time. One is interested in studying the details of
this coarsening process.

One can think of the model in analogy to a $q \rightarrow \infty$
state Potts system, where each site is assigned spin variable
$S=1,2\ldots q$.  In the initial configuration the spins in each stripe
are given the same state and the state associated with each stripe
increases in an ordered sequence $1,2,3 \ldots$ (see Fig.
\ref{fig:pict}(a)).  Each state $S=i$ propagates into states $S=j$
where $j>i$.  Clearly, the dynamics is such that the order parameter
(the density of particles of type $S$) is not conserved since the number
of particles of a given state changes in time. Related models
have previously been considered in the context of cyclic food chain
\cite{FKB} and the evolution of the spatial mosaic
of single-species domains was studied.

As a general motivation for the type of model we study, one may
think of the spin state as representing the height of a $D$
dimensional terraced surface. Thus the striped structure
corresponds to a sequence of terraces of increasing height; an
interface between two stripes corresponds to a step. The surface
evolves through particles being adsorbed or evaporated from the
steps. In this picture two coalescing interfaces correspond to a
terrace of a given height being eliminated from the system.
However, the model to be studied in this paper should be not
viewed as a microscopic description of this particular growth
process. Rather it provides a very simple dynamics that leads to
coarsening of stripes with no bubbles or overhangs in the interfaces between
stripes. In order to construct a realistic
microscopic model for a particular system (such as the terraced
surface) one would have to add other features that would
complicate analysis of the model.

The dynamical process described above is a generalization of the
reaction diffusion process $A+A \rightarrow A$, in which diffusing
$A$ particles (either with or without a drive) undergo a merging
reaction as they collide. In the present generalization the coalescing
objects are not particles but extended objects, i.e. the
interfaces between stripes, which are manifolds in $d=D-1$
dimensions.  The $A+A \rightarrow A$ reaction diffusion process
has been studied in detail over the years \cite{Redner}. It has
been found that for both biased and unbiased diffusion the density
(or equivalently the average distance between particles) decays as
$t^{-D/2}$ for $D<2$ and $t^{-1}$ for $D>2$ \cite{PCG}.  It is of
interest to investigate how this behavior is changed for the
generalization to coalescing manifolds.  In the present context
the manifolds themselves evolve in time, due to a roughening
process, which could also affect the behavior.

The scaling properties of driven manifolds are different from those of
stationary ones. It is of interest to consider the evolution of
striped structures when the interfaces are non-driven.  In this case
the dynamics can be related to an energy function. As we shall see,
the coarsening process is affected by whether or not the interfaces are
driven, unlike the case of coalescing particles described in
the previous paragraph.

In studying the model we find that the coarsening process may be well
understood by considering the scaling properties of an isolated
interface.  We find that the evolution is different at early and late
times. In both regimes the average domain size $\ell(t)$ grows
algebraically in time. The different regimes, however, are
characterized by different exponents. These exponents are determined
by the universal properties of the isolated interfaces and can
therefore be deduced using scaling arguments.  Moreover the exponents
depend on whether the interfaces are driven or not.

The paper is organized as follows: In Section 2 we define the
model. In Section 3 a scaling argument which is based on the
properties of an isolated interface is used to deduce the coarsening
behavior. Monte-Carlo simulations which verify the scaling argument
are described in Section 4. In Section 5 the special case where the
interfaces are non-driven and energy function exists is studied
using similar scaling arguments and Monte-Carlo simulations. We
conclude in Section 6 with a discussion.

%%%%%%%%%%%%%%%%%%%%%%%%%%%%%%%%%%%%%%%%%%%%%%%%%%%%%%%%%%%%%%%%%%%%%%%%%%%%%%%
\section{The coalescing interfaces model}
Our model is defined on a $D$-dimensional hyper-cubic lattice with
periodic boundary conditions. The lattice is infinite along one of
its axes, which we label $x$, and is of size $L$ along the $d$
other axes labeled ${\bf y} = (y_1,y_2,\ldots y_d)$. Each site
${\bf r}=(x,{\bf y})$ can be occupied by one of $q {\rightarrow}
\infty$ species of particles $S_{\bf r}=1,2, \ldots q$.  In the
initial configuration particles are arranged in stripes of single
species perpendicular to the $x$ direction.  The stripes are in
increasing order of particle type, their average extent along the
$x$ direction is $\ell_0$ and the interfaces between them are
flat. We are interested in constructing a model where species $i$
propagates into species $j$ when $j>i$. In view of the initial
configuration this can be simply achieved through the dynamics
we now describe.

The model evolves under random sequential dynamics according to
the following rules: at each time step two neighboring sites along
the $x$ direction ${\bf r}=(x,{\bf y}),{\bf r'}=(x+1,{\bf y})$ are
chosen randomly.  If the particles at sites ${\bf r}, {\bf r'}$
are not of the same species, then we proceed in one of the
following ways: with probability $1/(1+p)$, where $p<1$, the
update $S_{\bf r'}\to S_{\bf r}$ is made, but only if the
particles at ${\bf r}$ and its neighbors along the {\bf y}
directions are of the same species; with probability $p/(1+p)$ the
update $S_{\bf r}\to S_{\bf r'}$ is made, but only if the
particles at ${\bf r'}$ and its neighbors along the {\bf y}
directions are of the same species. For example, in $D=2$ the net
effect of this algorithm is for the following moves to occur with
the relative rates indicated above the arrows
\begin{eqnarray}
\begin{array}{ |l | l |}
\hline S \, &  \\ \hline S & S'
\\ \hline S & \\\hline
\end{array}
&\stackrel{1}{\longrightarrow}&
\begin{array}{ |l | l |}
\hline S \, &  \\ \hline S & S \,
\\ \hline S & \\\hline
\end{array}
\nonumber \\ \label{ratesKPZ}\\
\begin{array}{|l | l |}
\hline & S' \\ \hline S \, & S' \\ \hline &S' \\\hline
\end{array}
&\stackrel{p}{\longrightarrow}& \begin{array}{|l | l |} \hline
&S'
\\ \hline S'  & S' \\ \hline & S' \\\hline
\end{array}
\nonumber
\end{eqnarray}
where an empty square indicates that the move is attempted
irrespective of the type of particle at that site.

Given the initial conditions, this dynamics ensures that the species
index always increases to the right, along the $x$ direction.  This is
equivalent to stating that the interfaces are single valued and that
the order among the interfaces is preserved.  Furthermore, when $p <
1$ the dynamics is such that species $i$ is driven preferentially to
the right into neighboring species $j$ where $j>i$.  The case $p=1$
is special in that interfaces are not driven.  In the following we
consider $p<1$ and we discuss the special case $p=1$ towards the end
of the paper.  Finally note that in this dynamics the densities of the
various species are clearly not conserved.

Under this dynamics an isolated interface, evolves as in a restricted
solid on solid (RSOS) model, by which it is meant that changes in the
$x$ position of the interface as one moves in ${\bf y}$ directions are
at most of magnitude 1.  Moving interfaces of this kind belong
to the Kardar-Parisi-Zhang (KPZ) universality class.

The coarsening process for a system of many interfaces can be
understood in terms of the scaling properties of an isolated
interface.  These properties will be discussed in detail in the
following section.  However, at this point it is useful to outline
the general picture that emerges.  Starting from a dense
collection of flat interfaces, at early times interfaces meet and
coalesce due to the evolution of their width, $W$, which measures
the lateral extent of the interface. Thus $\ell$, the average
distance between neighboring interfaces, scales with the width of
an individual interface. So for early times, the dependence of
$\ell$ on time is governed by the time evolution of the width
In this regime the growth is independent of $L$ and is
determined by
an  exponent $\beta$
\begin{equation}
 \ell(t) \sim W \sim t^\beta\;\;\; \mbox{at early times.}
\end{equation}

Since we are considering
systems  of finite size $L$ in the directions transverse to
the drive, the width saturates to some final value
$W_{\mbox{sat}}(L)$. Therefore at late times, when the width of the
interfaces has already saturated, interfaces meet due to fluctuations,
$\Delta h$, in their center of mass position $h(t)$. That is,
\begin{equation}
 \ell(t)
\sim\Delta h \sim \frac{t^\gamma}{L^\phi}\;\;\; \mbox{at late times.}
\end{equation}
where $\gamma$ and $\phi$ are exponents
to be determined.
Notice the
algebraic evolution of $\ell$ in the two
regimes which contrasts with the logarithmic evolution of $\ell$
found in the model studied in \cite{YBEM}.

The proposed picture is valid only at time scales much larger than
that required for the process of coalescence of two interfaces.  We
find that at short times, the coalescence process takes in a finite
time independent of $L$. This is because neighboring interfaces touch
at a finite density of points.  On the other hand in the long time
regime the time scale for coalescence is found numerically to be
proportional to $L$. This has to be borne in mind when testing the
above scaling picture.

The exponents $\beta$, $\gamma$ and $\phi$ are determined by the
single interface behavior.  In the next section we introduce the
scaling analysis in a more quantitative way to obtain the exponents
$\beta$, $\gamma$ and $\phi$.  Specifically, for $p<1$ the interfaces
we consider belong to the KPZ universality
class, so the two exponents are given in terms of known KPZ
exponents. At the special point $p=1$ the interfaces belong to the
Edwards-Wilkinson (EW) universality class, and the two exponents are
changed accordingly.

%%%%%%%%%%%%%%%%%%%%%%%%%%%%%%%%%%%%%%%%%%%%%%%%%%%%%%%%%%%%%%%%%%%%%%%%%%%%%%%
\section{Scaling analysis}
\label{Sec:scaling} As argued in the previous section the
evolution of the system is governed by the width of an isolated
interface at short times.  On the other hand, at long times the
width saturates due to the finite lateral extent of the system
whereas the fluctuation of the center of mass (average $x$
position) keeps increasing. Thus at long times the center of mass
fluctuations govern the coarsening behavior.  We now define these
quantities more precisely and quantify their scaling behavior.

The width of an interface, $W$, is defined by
\begin{equation}
W^2=\left\langle \frac{1}{L^d} \sum_{\bf y} ( x_{\bf y} - h )^2
\right\rangle, \label{width}
\end{equation}
where $x_{\bf y}$ is the location of the interface along the $x$
direction and $h$ is the center of mass (average location) of the
interface $h=\sum_{\bf y} x_{\bf y}/L^d$. The angular brackets denote
an average over the dynamics, starting from the given initial
condition. The fluctuation in the average location of the interface
$\Delta h$ is defined through
\begin{equation}
\Delta h^2 = \left\langle  \left( h -
\left\langle h \right\rangle\right)^2 \right\rangle, \label{height}
\end{equation}
where as before angular brackets denote an average over the dynamics.

In general for a fluctuating
interface
the scaling behavior of the width is given
by (see for example \cite{Krug2})
\begin{equation}
W=t^{\beta} f\left( \frac{t}{L^z} \right),
\label{wscaling}
\end{equation}
where $z$ is the dynamic exponent.
The scaling function $f(x)$ is constant for $x \ll 1$ while $f(x) \sim
x^{-\beta}$ for $x \gg 1$. This implies,
\begin{equation}
W \sim \left\{ \begin{array}{l l}
 t^{\beta} \; , & {\rm for} \;\;\; t/L^z \ll 1 \\
 & \\
  L^\alpha \; , & {\rm for} \;\;\; t/L^z \gg 1.
\end{array} \right.
\label{wvalue}
\end{equation}
where $\alpha = \beta z$.

The fluctuation in the average location of the interface, $\Delta
h$, is dimensionally equivalent to the width so we may write, in
a similar fashion to (\ref{wscaling}),
\begin{equation}
\Delta h=t^{\beta} g\left( \frac{t}{L^z} \right).
\label{dhscaling}
\end{equation}
For long times we expect the center of mass of the interface
to diffuse, namely  we expect  $\Delta h \sim
t^{1/2}$, so that $g(x) \sim x^{1/2-\beta}$ for $x \gg 1$.
Thus
\begin{equation}
{\rm when} \quad t \gg L^z, \qquad \Delta h \sim \frac{t^{1/2}}{L^{\varphi}}
\label{dhscalinglt}
\end{equation}
where  $\varphi=z/2-\alpha$.

For short times one needs to consider the specific universality class
to which the interfaces belong. We start by restricting ourselves to
the case $p < 1$ (the case of $p=1$ will be discussed in Section
\ref{sec:EW}). In this case the dynamics is biased along the $x$
direction, and the interfaces clearly belong to the KPZ universality
class.

For KPZ interfaces at short times $t \ll L^z$ it has been argued
\cite{Krug1} that there are two regimes depending on the ratio
$(d+4)/z$. That is,
\begin{equation}
{\rm when} \quad t \ll L^z, \qquad\Delta h \sim \left\{ \begin{array}{l l}
{\displaystyle \frac{t^{\theta}}{L^{d/2}}} \; , & {\rm for} \;\;\; d+4 < 4z \\
& \\
{\displaystyle \frac{t}{L^{2(z-1)}}} & {\rm for} \;\;\; d+4 >4z \\
\end{array} \right.
\label{dhvalueKPZ}
\end{equation}
where $\theta=(d+4)/2z-1$.

We now use the width of the interface, $W$, and the
fluctuation in its average position, $\Delta h$, to deduce the
average distance between the interfaces as a function of time
$\ell (t)$. As stated above we will show that $W$ controls
the early time behavior while $\Delta h$ controls the late time
behavior of the system. We assume that initially the average
distance between the interfaces $\ell_0$ is much smaller than the
width of a fully developed isolated interface $L^\alpha$. It will
be shown that two distinct coarsening behaviors are expected for $t
\ll L^z$ and $t \gg L^z$.

Consider first the long time ($t \gg L^z$) behavior of the system. From
Eqs. \ref{wvalue} and \ref{dhvalueKPZ} one can see that $\Delta h
\gg W$ and therefore $\Delta h$ is expected to control the time scale
on which interfaces coalesce. This implies that the average distance
between interfaces should behave as
\begin{equation}
\ell \sim  \frac{t^{1/2}}{L^{\varphi}} \;\;\;\;\;\; t \gg L^z.
\label{ltKPZ}
\end{equation}
Next, consider the short time behavior ($t \ll L^z$). In this case
it is  straightforward to verify, using the hyper-scaling relation
$\alpha+z=2$,  that $W \gg \Delta h$ for
both regimes in Eq. \ref{dhvalueKPZ}.
Thus at short times the coarsening behavior is controlled by the
width of the interface and is  expected to behave as
\begin{equation}
\ell \sim t^{\beta}  \;\;\;\;\;\; t \ll L^z. \label{stKPZ}
\end{equation}

We have thus shown that $\Delta h$ controls the long time behavior
and $W$ the short time behavior. The short time behavior is
expected to be lost if the initial spacing between the interfaces
is larger than $W_{\mbox{sat}} \sim L^\alpha$. Also, in
models for which the interfaces are smooth ($\alpha \leq  0$) the
short time behavior is not expected to be seen.

In the following
we verify the predictions of the scaling analysis by numerical
studies of the models in $D=2$ and $D=3$ dimensions.

%%%%%%%%%%%%%%%%%%%%%%%%%%%%%%%%%%%%%%%%%%%%%%%%%%%%%%%%%%%%%%%%%%%%%%%%%%%%%%%
\section{Monte Carlo simulations}
We first consider the model in $D=2$ dimensions. The model is
simulated on a lattice of size $L_x \times L_y$ where $L_y$ gives
the lateral size denoted by $L$ in the previous sections.
Periodic boundary conditions are used both in the $x$ and $y$
directions.   Particles of
different types are initially ordered in stripes parallel to the
$y$ axis. The positions of the interfaces between the stripes are
chosen randomly, such that the mean distance between them is
$\ell_0$. At each step two neighboring sites in the $x$ direction,
which are occupied by different species are chosen randomly, and a
move is made according to Eq. \ref{ratesKPZ} where we have used
$p=0$ in our simulations. In this way we maximize the speed of the
simulation. After each such move, time is advanced by
$\tau=1/\epsilon_x$, $\epsilon_x$ being the total number of
nearest neighbor pairs in the $x$ direction occupied by different
species. The algorithm would be equivalent to a usual Monte Carlo
simulation if $\tau$ were drawn from a Poisson distribution  with
mean $1/\epsilon_x$. Here we make the approximation
$\tau=1/\epsilon_x$ which is valid as long as $\epsilon_x$ is
large.

\begin{figure}
\begin{center}
\epsfxsize 16 cm %\epsfysize 3 cm
\epsfbox{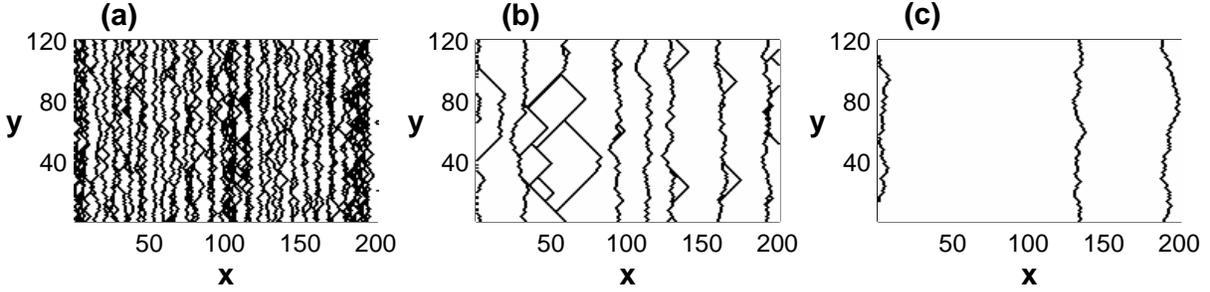} \caption{Snapshots of a $d=1$ system at
(a) $t=10$,  (b) $t=10^3$ and (c) $t=10^5$. Here $L_x=1000$,
$L_y=120$
  and the average initial spacing between the interfaces is $\ell_0=4/3$
  lattice sites. For clarity only quarter of the system in the $\widehat{x}$
  direction is presented.}
  \label{fig:KPZconfig}
\end{center}
\end{figure}

In Fig.  \ref{fig:KPZconfig} typical snapshots of the system at early,
intermediate and late times are presented for a system of size
$L_y=120$ and $L_x=2000$. The initial spacing between the interfaces
is $\ell_0=4/3$.  In Fig. \ref{fig:KPZconfig}(a)  the
early time behavior is shown. Here interfaces meet each other at
several points indicating that their width is of the same order as the
spacing between them. Note that in order to see such configurations
the initial spacing between the interfaces must be much smaller than
the final width $W_{\mbox{sat}}$ of an isolated interface.
At the short time regime the mean domain
size $\ell$ is expected to scale according to Eq. \ref{stKPZ} which
for $d=1$ reads $\ell \sim t^{1/3}$.

At late times (Fig. \ref{fig:KPZconfig}(c)) the distance between
interfaces is much larger than their width, and interfaces meet
due to fluctuations in their center of mass location. In this
regime $\ell$ is expected to scale according to Eq.(\ref{ltKPZ}),
which in the $d=1$ case is just $\ell \sim t^{1/2}/L^{1/4}$.

In the intermediate regime $\Delta h$ is comparable with $W$ and we
expect a crossover from the early to late time behavior.
Fig. \ref{fig:KPZconfig}(b) shows a typical configuration at this
stage of the dynamics.  We see from the figure that in this regime
there is a significant probability of three interfaces meeting at a
point. This results in some arrested local configurations where an
interface is temporarily frozen in locally triangular forms and the
apparent width of the interface becomes large. These configurations
are then released when a fourth interface approaches from the left and
sweeps through the triangular forms.  This effect is a direct
consequence of the RSOS condition of the model.  We have explored
variations of the model in which the RSOS condition is relaxed, and
found a variety of model specific effects in this intermediate
regime. Nevertheless the early and late time behaviors are not
affected by the details of the model.
\begin{figure}
\begin{center}
\epsfxsize 8 cm
\epsfbox{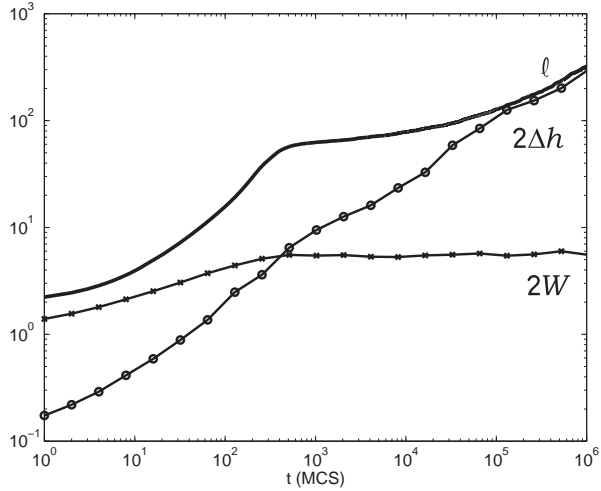}
 \caption{Results of simulations for $D=2$.
  The mean spacing between interfaces $\ell$ as a
  function of time is shown, along with twice the width $W$
  and twice $\Delta h$, the fluctuations in average location  of a single
  isolated interface. Note that at late time $\ell$
  coincides with $2 \Delta h$ as expected while at early times $\ell$
  is parallel to $W$. Here $L_x=1000$, $L=120$, and time is measured
  in Monte Carlo sweeps. The results are obtained from
  an average over 20 simulations for $\ell$ and 200 simulations for
  $W$ and $\Delta h$.
  }
  \label{fig:KPZcoarsen}
\end{center}
\end{figure}

The overall picture we have outlined is quantified in Fig.
\ref{fig:KPZcoarsen}. In this figure the average distance between
stripes $\ell(t)$ is plotted as a function of time and is compared
with the width $W$ and fluctuation in the average interface location
$\Delta h$ of a single isolated interface.  We calculate $\ell$
through $\ell=L_xL_y/\epsilon_x$ where $\epsilon_x$ is defined as
above, namely, the total length in the $y$ direction of the
interfaces.  Similar results were obtained for various system sizes.

In Fig. \ref{fig:KPZcoarsen} one can see that at early time indeed
the behavior of $\ell$ follows closely that of $2W$, while at late
times it follows closely that of $2\Delta h$. The factors of 2 are
because the growth of $\ell$ is actually controlled by the sum of
the fluctuations of two neighboring interfaces.  In the
intermediate regime where $\Delta h$ and $W$ are comparable $\ell
$ deviates from both curves. This reflects the triangular
configurations observed in this regime as discussed before.

\begin{figure}
\begin{center}
\epsfxsize 8 cm
\epsfbox{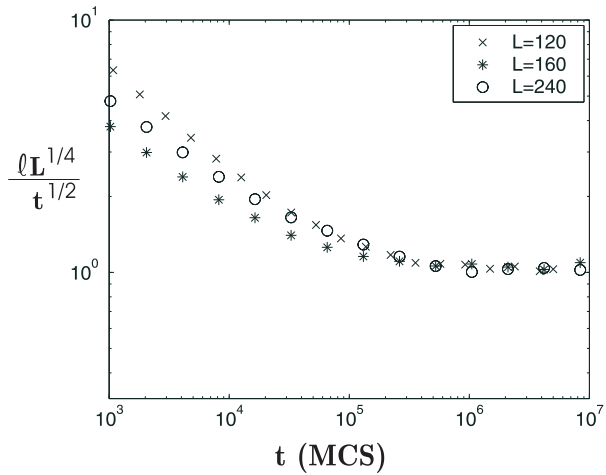} \caption{Scaled average domain size $\ell
L^{1/4} / t^{1/2}$  for systems of size $L =
120,160,240$. The collapse at long times demonstrates the scaling
predicted by Eq. \ref{ltKPZ}.} \label{fig:ellltKPZ}
\end{center}
\end{figure}

We now make more precise tests of the scaling predictions.  At late
times ($t \gg L^z$) the scaling argument predicts a specific
dependence of $\ell$ on $t$ and $L$, $\ell \sim t^{1/2}/L^{1/4}$.  To
test this, in Fig. \ref{fig:ellltKPZ} we plot $\ell L^{1/4}/t^{1/2}$
against $t$ for various system sizes. One can see that at late time
indeed the data collapses confirming the predicted behavior.

To get a more quantitative test of the predictions of the scaling
argument at early times ($t \ll L^z$) is more difficult.  The reason
is that one has to access a regime in which $\Delta h \ll W$ while, at
the same time, $W$ displays its early time growth behavior. To obtain
a small $\Delta h$, and to increase the time window in which $W$ keeps
growing, we require a large lateral size $L$.  Also note that at short
times a coalescence process of two interfaces occurs in a finite time
which does not grow with $L$. This is because in this regime the
neighboring interfaces touch at a finite density of points.

\begin{figure}
\begin{center}
\epsfxsize 8 cm
\epsfbox{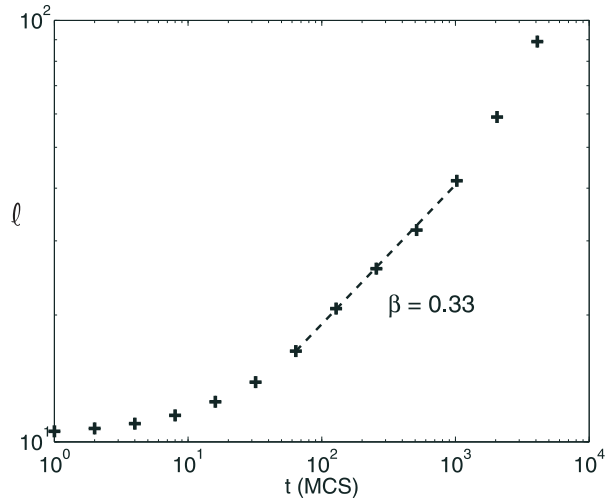} \caption{Average domain size $\ell$ as a
function of time, for a $D=2$ system with $L=10^4$. At short time
one can see $\ell \sim t^{\beta}$ with $\beta = 0.33 \pm 0.01$
(dotted line).} \label{fig:ellstKPZ}
\end{center}
\end{figure}

In Fig. \ref{fig:ellstKPZ}
the average distance between the domains
$\ell$ is plotted as a function of $t$ for a system with $L=10^4$ and
$\ell_0=10$. After a transient time the domain size $\ell \sim
t^{\beta}$ over more than a decade, where $\beta = 0.33 \pm 0.01$
as expected.

\begin{figure}
\begin{center}
\epsfxsize 8 cm
\epsfbox{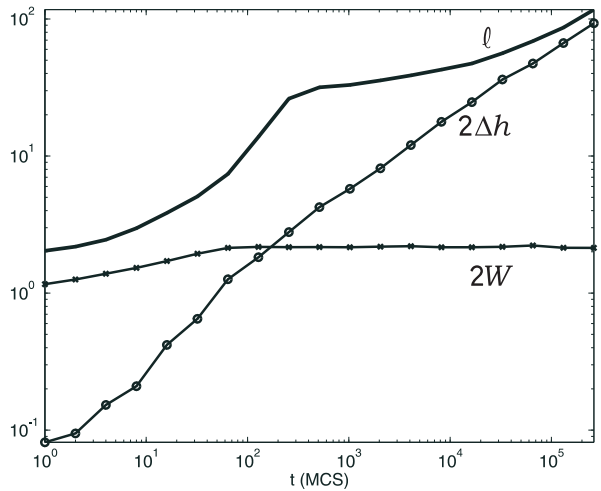} \caption{Results of simulations for $D=3$.  The
  mean spacing between interfaces $\ell$ as a function of time is
  shown, along with the width $W$ and the fluctuations in average
  location $\Delta h$ of a single isolated interface. Here $L_x=1000$,
  $L=L_y=L_z=20$, and time is measured in Monte Carlo sweeps. The
  results for $\ell$ are obtained from a single run, while those for
  $W$ and $\Delta h$ are averaged over 100 runs.}
  \label{fig:KPZ2d}
\end{center}
\end{figure}

Next we present results obtained from simulation of a three
dimensional system ($D=3$). In Fig. \ref{fig:KPZ2d} the average distance
between interfaces $\ell$ is plotted along with $W$ and
$\Delta h$ for a system with $L_y=L_z=20$. Again one can see that
at early times $\ell$ follows closely $W$ while at late times
it follows $\Delta h$. While these simulations confirm
the general picture outlined above, quantitative estimates
of the exponents would require a more elaborate study.

%%%%%%%%%%%%%%%%%%%%%%%%%%%%%%%%%%%%%%%%%%%%%%%%%%%%%%%%%%%%%%%%%%%%%%%%%%%%%%%
\section{Unbiased dynamics}
\label{sec:EW}

We turn now to study the case $p=1$. In this case one can associate
with the dynamics a well-defined local energy function.  Since at each
update the number of $x$ neighbor pairs of different species is
reduced, the energy is Potts-like in the $x$ direction with unlike
neighbor pairs costing energy.  Unlike pairs in the $y$ direction do
not cost energy provided that the RSOS condition is satisfied.  Any
step of the dynamics carried out, serves to lower the energy. In this
way the dynamics can be thought of as a zero temperature system
approaching an equilibrium state.

Single interfaces in this special case belong to the EW universality
class, rather than to the KPZ class. Therefore the scaling analysis of
Section \ref{Sec:scaling} has to be applied to  EW type interfaces. This
modifies the values of the exponents and also the short time behavior
of $\Delta h$.

For interfaces belonging to the EW universality class $\Delta h
\sim t^{1/2}$ for all times.
Therefore the scaling form
(\ref{dhscaling}) implies
\begin{equation}
\Delta h \sim \frac{t^{1/2}}{L^{\phi}} \label{dhEW}
\end{equation}
for all times, with $\phi = z/2-\alpha$ as before.
We use the exact values of the EW exponents:
$z=2$, $\alpha=1-d/2$, $\beta= 1/2-d/4$ to find
\begin{equation}
\Delta h \sim \frac{t^{1/2}}{L^{d/2}}\;.
\end{equation}

As in the previous case one can now determine
the behavior of
$\ell(t)$. For long times ($t \gg L^2$) using (\ref{wvalue}),
(\ref{dhEW}) and the values of the EW exponents one finds that
$\Delta h \gg W$ and $\Delta h$ is expected to control the mean distance
between interfaces. Therefore
\begin{equation}
\ell \sim \frac{t^{1/2}}{L^{d/2}}\qquad t \gg L^2\;.
\end{equation}

For short times ($t \ll L^2$)
we have
$W \sim t^\beta \gg \Delta h$.
Therefore the coarsening is governed by the width of the
interfaces and
\begin{equation}
\ell \sim t^{1/2-d/4} \qquad t \ll L^2. \label{stEW}
\end{equation}

To summarize, in this case, just as in the $p \neq 1$ case,
$\Delta h$ controls the long time behavior and $W$ the short
time behavior.  The difference between the two cases is due to the
different values of the exponents $\alpha$, $\beta$ and $z$ and is thus a
direct consequence of the universality class.

Monte-Carlo simulations in $D=2$ with no bias support these
findings. The simulations show that indeed at early times $\ell \sim
t^{1/4}$ and at late times $\ell \sim t^{1/2}/L^{1/2}$ in agreement
with the scaling argument and the known EW exponents. The results for
the evolution of $\ell$ are qualitatively similar to those of the KPZ
case and we do not display them here.

\section{Discussion}
In this work we have considered the evolution of a class of striped
structures. A coarsening process takes place in which interfaces
separating the stripes meet and coalesce.  The dynamics may thus be
viewed as a generalization of the particle reaction diffusion process
$A+A \to A$ to extended objects, the interfaces. In this generalization
the coalescing objects have internal structure that in turn affects
the evolution. Thus one may have different scaling laws depending on
the properties of the coalescing objects.

We studied a microscopic model defined through the dynamics of the
interfaces that could realize both driven and undriven interfaces.
Both cases lead to rough interfaces along with the average stripe
width growing as a power law in time.  However, the two cases lead
to different scaling exponents.  In the driven case an isolated
interface exhibits KPZ behavior, while in the undriven case EW
behavior is obtained.  It is this scaling behavior of an isolated
interface that determines both short and long time coarsening
regimes of the stripes.
We have shown that at early time the
coarsening dynamics is determined by the width of the interfaces,
while at short time it is determined by the fluctuations in the
locations of the interfaces. The behavior is different from that
of other driven models where the interfaces are smooth and the
coarsening is logarithmically slow in time \cite{YBEM}.

The analysis performed in this paper should apply also to other
classes of interfaces. For example, the scaling behavior of moving
fronts or interfaces has been studied for the case of `pulled'
fronts, moving into an unstable phase. Recently, it has been
suggested that such an interface is not of the KPZ
type \cite{RDbA,TvS,REvS}. This indicates that there might be other
possible types of scaling behavior than those we have studied in
this paper.  It would be of interest to explore such possible
dynamics and study the resulting coarsening.
\vspace*{2em}

\noindent {\bf Acknowledgments}: MRE and DM thank the Einstein Center
and the University of Edinburgh
for support and hospitality during mutual visits;
the support of the Israeli Science Foundation is gratefully acknowledged.

\end{document}